# Electron-phonon coupling in *d*-electron solids: A temperature-dependent study of rutile TiO₂ by first-principles theory and two-photon photoemission

Honghui Shang [1,*] Adam Argondizzo,[2] Shijing Tan,[2] Jin Zhao,[3] Patrick Rinke,[4]
Christian Carbogno,[1] Matthias Scheffler,[1] and Hrvoje Petek [2,†]

[1]*Fritz-Haber-Institut der Max-Planck-Gesellschaft, Faradayweg 46, D-14195 Berlin-Dahlem, Germany*
[2]*Department of Physics and Astronomy and Pittsburgh Quantum Institute, University of Pittsburgh, Pittsburgh, Pennsylvania 15260, USA*
[3]*Department of Physics, University of Science and Technology of China, Hefei 230026, China*
[4]*Department of Applied Physics, Aalto University, P.O. Box 11100, FI-00076 Aalto, Finland*

Rutile TiO₂ is a paradigmatic transition-metal oxide with applications in optics, electronics, photocatalysis, etc., that are subject to pervasive electron-phonon interaction. To understand how energies of its electronic bands, and in general semiconductors or metals where the frontier orbitals have a strong *d*-band character, depend on temperature, we perform a comprehensive theoretical and experimental study of the effects of electron-phonon (*e-p*) interactions. In a two-photon photoemission (2PP) spectroscopy study we observe an unusual temperature dependence of electronic band energies within the conduction band of reduced rutile TiO₂, which is contrary to the well-understood *sp*-band semiconductors and points to a so far unexplained dichotomy in how the *e-p* interactions affect differently the materials where the frontier orbitals are derived from the *sp*- and *d* orbitals. To develop a broadly applicable model, we employ state-of-the-art first-principles calculations that explain how phonons promote interactions between the Ti-3*d* orbitals of the conduction band within the octahedral crystal field. The characteristic difference in *e-p* interactions experienced by the Ti-3*d* orbitals of rutile TiO₂ crystal lattice are contrasted with the more familiar behavior of the Si-2*s* orbitals of stishovite SiO₂ polymorph, in which the frontier 2*s* orbital experiences a similar crystal field with the opposite effect. The findings of this analysis of how *e-p* interactions affect the *d*- and *sp*-orbital derived bands can be generally applied to related materials in a crystal field. The calculated temperature dependence of *d*-orbital derived band energies agrees well with and explains the temperature-dependent inter-*d*-band transitions recorded in 2PP spectroscopy of TiO₂. The general understanding of how *e-p* interactions affect *d*-orbital derived bands is likely to impact the understanding of temperature-dependent properties of highly correlated materials.

## I. INTRODUCTION

Electron-phonon (*e-p*) interaction pervasively impacts crystalline materials, is the key physical property that defines conductivity, superconductivity [1,2], and solar energy conversion [3–6]; moreover, it determines the temperature scales for quantum phase transitions [7,8] and electron-spin-dependent phenomena [9], controls reactivity of catalytic chemical reactions [10,11], influences the topological properties of high-*Z* materials, defines the nonradiative dynamical processes of photogenerated carriers, etc. Despite importance of *e-p* interaction in the electronic, optical, chemical, and thermal properties of materials, a broadly applicable and simple to apply theory or physical principles that consider the atomic character of the frontier orbitals, and based on that can explain the broad spectrum of *e-p* interactions, does not exist. Such a theory is of high current topical interest because technological advances employing *sp*-orbital-based semiconductors are reaching maturity, and high-*Z* materials with *d*- or *f*-orbital-based frontier bands are of increasing interest as advanced semiconductors [12], topological materials exhibiting the quantum spin-Hall and Rashba effects [13–15], 2D materials undergoing ultrafast charge, spin, or exciton transfer [3,16], energy-harvesting topological materials with strong crystal-field effects on transport [17], etc. How optical excitation modulates the *e-p* interaction and how phonon excitation affects quasiparticle dynamics is likely to play an increasing role in understanding and controlling the ultrafast quasiparticle correlation in solids [3,4,16,18–23].

The conventional electronic materials are considered to have normal manifestation of *e-p* interaction in that properties such as their fundamental band-gap energy decreases with temperature (Varshni effect [24]), while others are considered to be abnormal, because the opposite is observed (inverse Varshni effect [25]). In this paper, we demonstrate that such opposing behaviors can be simply manifestations of their frontier orbital symmetries: the degenerate electronic orbitals of an atom are split by the crystal field of the ionic lattice,

*State Key Laboratory of Computer Architecture, Institute of Computing Technology, Chinese Academy of Sciences, Beijing 100190, China.

†Corresponding author: petek@pitt.edu

and the relative motions of lattice ions, as described by the normal modes of vibration, can either stabilize or destabilize them. We show that the *e-p* induced electronic band renormalization strongly depends on the atomic orbital symmetries and whether the normal modes of thermally excited vibrations enhance the attractive or repulsive interactions with the neighboring ions. Thus, a general understanding of how the *e-p* interaction impacts the electronic and quantum behavior of materials must consider the fundamentals of how the atomic orbitals that form the frontier electronic bands are affected by the crystal-field symmetry and motion of the surrounding atoms that define their electronic environment.

Our study of *e-p* interaction is cast around rutile TiO$_2$, a large band-gap semiconductor, which is abundant and versatile, and has pervasive applications in optics, electronics, photocatalysis, as a white pigment, a food additive, and a tractable model for more complex metal-oxide materials [26–29], that derive their properties from the related photon-electron, electron-electron (*e-e*), and *e-p* interactions. TiO$_2$ is a suitable model because it has been extensively studied [26,29–32] and its frontier valence electronic orbitals derive from the *sp* orbitals of O$^{2-}$ ligands, while its conduction band derives from the *d* orbitals of Ti$^{4+}$ metal ions; the octahedral coordination of the ligands forms a crystal field that defines the band formation and is common to more complex metal oxides. *e-p* Interactions in the common electronic semiconductors and metals are well understood, but this knowledge may not be germane to transition-metal oxides because their valence- and conduction bands are derived from different atomic orbitals [25,33–35]. Although optical spectroscopy and electronic transport of TiO$_2$ have burgeoned into large and active research fields [36–39], a systematic understanding anchored by fundamental principles of how *e-p* interaction affects these properties is not available, but is requisite for advancing applications of TiO$_2$, as well as those of other materials where the ligand coordination of *d*-electron orbitals is a key to exotic electronic properties [13,15,17]. Although spectroscopic studies of metal-oxide optical properties are few, recent measurements on ZnO [40] show complex dynamics that depend strongly on the nature of the frontier orbitals, e.g.,Ti-3*d* vs Zn-4*s*. *e-p* Interactions are most vividly manifested in temperature dependence of the optical properties of solid-state materials, and thus constitute a well-defined target for incisive comparisons between experiment and theory. In addition to the optical properties, *e-p* interactions also affect the carrier transport, which is crucially important for electron correlation and photocatalysis, but is also difficult to relate to the specific electronic structures of materials [41–43]. Here we focus on the temperature dependence of the electronic bands, from the first-principles understanding of *e-p* interactions that affect the relevant atomic orbitals, and how they are manifested in the optical properties of materials.

Our experimental study by two-photon photoemission (2PP) spectroscopy of reduced TiO$_2$ samples motivates the theoretical study of the *e-p* interactions of *d* bands of solids [44]. TiO$_2$ is an excellent model, because its valence- and conduction bands are defined, respectively, by the O-2*p* and Ti-3*d* orbitals of $t_{2g}$ symmetry [30]. For an octahedrally coordinated ion, the ligand 2*p* orbitals have a bonding interaction with the central ion, but the lowest $t_{2g}$ orbitals have a weak

$\pi*$ antibonding character and higher-energy $e_g$ orbitals have a stronger $\sigma*$ antibonding character [45]. Thus, the symmetry-forbidden $t_{1g}$-$t_{2g}$ band-gap transition in TiO$_2$ promotes an electron from a predominantly O-2*p* bonding orbital at the valence-band maximum (VBM) to a weakly antibonding $t_{2g}$ Ti-3*d* orbital of the conduction band (CB) [45,46]. These characteristics make TiO$_2$ interesting and broadly relevant to discussion of how the *e-p* interactions affect the electronic bands that differ in occupations of the *d* orbitals. The forbidden nature of the fundamental band gap of TiO$_2$, however, makes it difficult to probe experimentally the *e-p* interactions by optical methods. Instead, we perform 2PP spectroscopy on the $t_{2g}$-$e_g$ transition of reduced TiO$_2$ samples [47,48], where thermal processing creates O-atom vacancy defects, thereby transferring excess electrons to a Ti-3*d* defect state 0.9 eV [49] below the Fermi level, $E_F$.

## II. THEORETICAL FRAMEWORK

First, we elaborate general aspects of *e-p* interactions in solids. Within the harmonic approximation, the temperature renormalization of the electronic eigenstates in a solid is given by [34]

$$
\begin{aligned}
\delta\varepsilon_{nk}(T) &= \int dq \sum_j \frac{\partial \varepsilon_{nk}}{\partial n_{jq}(T)}\left[n_{jq}(T)+\frac{1}{2}\right] \\
&= \sum_q w_q \sum_j \frac{\partial \varepsilon_{nk}}{\partial n_{jq}(T)}\left[n_{jq}(T)+\frac{1}{2}\right] \\
&= \sum_{jq} f_{jq}\left[n_{jq}(T)+\frac{1}{2}\right],
\end{aligned}
\tag{1}
$$

with

$$
\frac{\partial \varepsilon_{nk}}{\partial n_{jq}(T)} = \frac{\hbar}{2\varpi_{jq}M_{jq}}\frac{\partial^2 \varepsilon_{nk}}{\partial z_{jq}^2},
\tag{2}
$$

where $jq$ and $nk$ are the phonon and electronic band indices and wave vectors. $z_{jq}$ Is the collective atomic displacement along the vibrational eigenmode $jq$. Here $w_q$ is weight of the reciprocal-space $q$ point in the integration. The Bose-Einstein distribution $n_{jq}(T) = \frac{1}{\exp(\hbar\varpi_{jq}/k_BT)-1}$ gives the temperature-dependent quantum-mechanical phonon occupation numbers that account for the thermal effects, where $\varpi$ refers to the phonon frequency. Of primary interest are $f_{jq} = w_q\frac{\partial \varepsilon_{nk}}{\partial n_{jq}(T)}$, i.e., the weighted *e-p* factors, which describe the magnitude and sign of the *e-p* coupling (phonon mode-specific band renormalization), both the phonon occupation number and the *e-p* factors contribute to the final temperature dependent curves; here we use them to determine the temperature-dependent phonon contributions to band shifts of TiO$_2$ and SiO$_2$.

As shown in Eq. (2), the key ingredient to the temperature renormalization of the electronic eigenstates is the second-order derivative of eigenvalues with respect to the atomic displacements. This quantity can be calculated either with a nonperturbative frozen-phonon approach [50–52] or with perturbative analytical methods based on density-functional perturbation theory (DFPT) [53–56]. In the perturbative framework, it has two leading terms [34,57–59]: the Fan term, which accounts for the first-order phonon-induced perturbation of the electronic energy up to second order in the ionic displacements and the Debye-Waller (DW) term, which is the second-order phonon-induced perturbation of the Hamiltonian, with diagonal components that can be recast in terms of first-order matrix elements in the rigid-ion approximation. For this reason, only the diagonal components of the DW term are easily accessible in DFPT, whereas the frozen-phonon approach also naturally captures the (often negligible) off-diagonal ones [51,56,60]. Here, we use the frozen-phonon approach that we previously also applied to polarons in ZnO [61]. Then we add the polar correction proposed by Nery and Allen [62] beyond the frozen-phonon approach, which has negligible effects on qualitative trends discussed here [63].

Second, temperature dependence of the electronic bands has, in addition to the contribution from atomic motion, a contribution from the thermal expansion/contraction of the lattice that is caused by the anharmonic terms in the lattice potential [64,65]. To compute the thermal expansion/contraction, the quasiharmonic approximation is used [65]. We minimize the free energy with respect to volume at a given temperature to obtain the temperature-dependent equilibrium volume. In this work, we primarily focus on the $e$-$p$ coupling-related effects, i.e., the atomic contributions given by Eq. (1). The corrections stemming from the thermal expansion are later introduced to make a quantitative comparison with the experimental data.

Third, we consider how the crystal field affects electronic orbitals of different symmetry. The temperature dependence of semiconductor electronic band gaps has been investigated for a large group of materials, which generally show the "normal" bathochromic behavior (redshift), i.e., the band gaps decrease as the temperature increases [24,66]. Common materials that follow this trend are the low-Z Group IV and III–V semiconductors, whose electronic properties derive from covalent bonds involving the $s$- and $p$ orbitals of the constituent materials [67–69]. Because such materials are commonly used in electronic applications, their temperature behavior can color our expectations of what constitutes the normal material behavior. Some materials, such as $CuGaS_2$ and $AgGaS_2$ [25], however, have an increasing gap at low temperatures (hypsochromic, blueshift), but exhibit the opposite bathochromic behavior at higher temperatures. Other materials, like copper halides [70], black phosphorus [71], and methylammonium lead iodide perovskite [72] exhibit a hypsochromic shift exclusively. Such properties, however, have been documented mainly experimentally [66], and the "anomalous" effects of temperature on the electronic band structures of some of these materials have been attributed to presence of noble metals with $d$ electrons [25]. Although a role of the $d$ orbitals in the $e$-$p$ coupling has been implicated, no *ab initio* explanation (or confutation) of this argument has hitherto been proposed, although Bhosale *et al.* [25] expressed an urgent need for such a theory. That the $d$ orbitals are responsible for the anomalous $e$-$p$ interactions is supported by hypsochromic shifts upon transient population of the Fe-3$d$ conduction band states that have been reported in the x-ray edge spectra of $Fe_2O_3$ [73–75]. Furthermore, the temperature or laser-induced metal-insulator phase transition of $VO_2$ provides indisputable evidence that $e$-$p$ interactions involving the $d$ orbitals can have profound effects on the electronic correlation properties of materials [76–79].

The seemingly anomalous shifts of states that involve $d$ orbitals can be understood qualitatively from the crystal-field theory [80], which describes how the energies of degenerate atomic orbitals tune when transposed from an isotropic field to a crystal field of a particular symmetry. According to the crystal-field theory, an exact octahedral or a tetrahedral field causes the five-degenerate $d$ orbitals of an atom split into $t_{2g}$ and $e_g$ symmetry subsets, respectively, with the $d_{xy}$, $d_{yx}$, and $d_{zx}$ in the former and $d_{z^2}$ and $d_{x^2-y^2}$ in the latter. An octahedral field shifts the $t_{2g}$ orbitals downward in energy and the $e_g$ orbitals upward in energy, while a tetrahedral field has the opposite effect. One should be aware that the symmetries may not be exact because of distortion of the atomic positions from high symmetries on account of chemical bonding. In rutile $TiO_2$ there is some mixing between orbitals associated with the $t_{2g}$ and $e_g$ symmetries, but that is included in our theory [32,47,81]. In typical materials, such as transition-metal oxides, the $t_{2g}$-$e_g$ splitting is on the order of eV. Thus, the phonon modulation of the crystal field and thermal change in the crystal unit-cell dimensions can strongly modulate energies of the $t_{2g}$ and $e_g$ orbitals differently from materials with $s$- or $p$-orbital-derived bands; this makes the crystal-field theory a natural, qualitative starting model for understanding the seemingly anomalous $e$-$p$ interactions of $d$-orbital-derived bands. An *ab initio* study of the temperature dependence of the $d$ bands can broaden our understanding of $e$-$p$ interactions in solid-state materials involving the transition and noble metals, and explain their seemingly anomalous and potentially useful behaviors.

Our experimental study of the temperature dependence of the electronic bands of $TiO_2$ by 2PP spectroscopy of samples motivates the theoretical study of the $e$-$p$ interactions of $d$ bands of solids [44]. Because $TiO_2$ is a wide-band semiconductor, unless the sample is reduced, it charges up during the photoemission measurements.

## III. EXPERIMENT

Besides the intrinsic chemical bonding, the electronic structure of $TiO_2$ can further be impacted by reduction. This typically occurs in the process of preparing a clean, crystalline surface under ultrahigh-vacuum conditions by thermal processes that remove O atoms from the lattice, such as by heating the sample above 800 K or ion bombardment. O atoms are removed in the form of $O_2$ molecules; this leaves two extra electrons per O atom in a $t_{2g}$ defect band with a maximum density at ∼0.9 eV below the $E_F$ [82]. The broad defect density extends up to $E_F$, which is ∼0.1 eV below the conduction-band minimum (CBM) [41,42,47,48]. Whether these electrons remain localized at $Ti^{3+}$ sites where the Ti-O bonds are broken, or are localized near the vacancies in polaron states, is under debate [42,83–85]. Because the theoretical localization energies are not very large, it is likely that excess electrons occupy multiple low-lying states at finite temperatures [82]. Experiments, like 2PP, require a conductive $TiO_2$ sample, and therefore have been performed on reduced $TiO_2$ samples, whereas theoretical calculations for describing the $e$-$p$ interactions in the conduction band are most conveniently performed for the stoichiometric $TiO_2$. Therefore, when comparing experiment and theory, this distinction should be understood, but we believe that the comparison is meaningful because both types of initial states belong to the $t_{2g}$ symmetry [81].

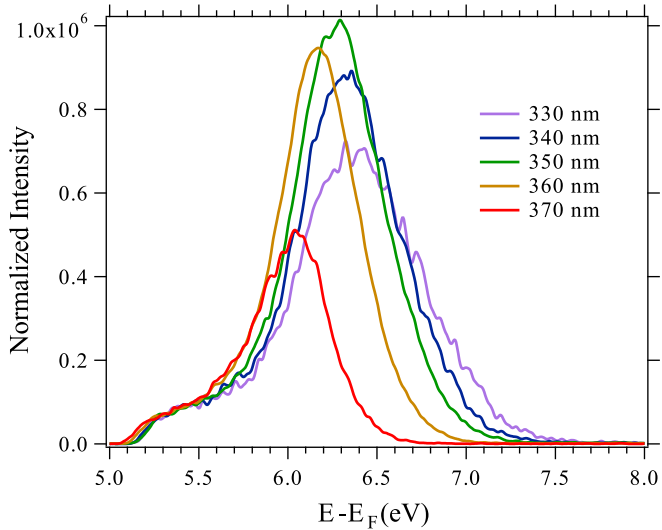

FIG. 1. 2PP spectra of $TiO_2$ taken with UV excitation between 370 and 330 nm with $p$-polarized light. The maximum intensity at 3.44–3.54 eV (350–360 nm) corresponds to the $t_{2g}$-$e_g$ resonance at 300 K for the sample aligned with the $[1\bar{1}0]$ axis in the optical plane.

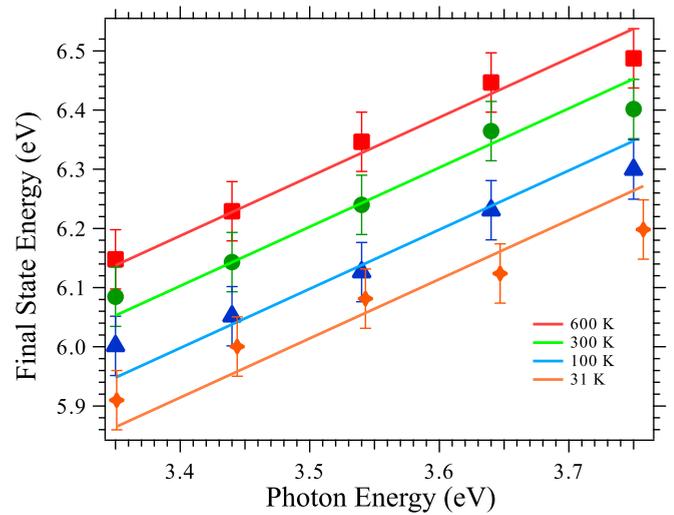

FIG. 2. Wavelength-dependent 2PP measurements for 600, 300, 100, and 31 K showing the position of the final photoelectron energy at which the $e_g$ state is observed vs the photon energy. The intermediate $e_g$ state energy decreases with temperature. A slope of 1 is fixed in fitting the photon energy dependence of the 2PP peak energies for all temperatures; the intercept of each plot gives the $e_g$-state energy relative to $E_F$. The data are obtained from $s$-polarized light for the [001] aligned sample.

UV-light excited 2PP spectra of reduced $TiO_2(110)$ single crystal can be enhanced by a single-photon resonance involving the $t_{2g}$-$e_g$ transition between the $TiO_2$-3$d$ bands [47,48]. The nominally $t_{2g}$-$e_g$ transition, which is resonant at ~3.6 eV at 300 K, occurs from initial state with the dominant $d_{xy} + d_{z^2}$ orbital character to two intermediate states with the dominant $d_{z^2}$ and $d_{xz} + d_{yz}$ orbital character [47,81] at 2.60 ± 0.05 and 2.73 ± 0.05 eV above $E_F$. Because of their symmetries, the $t_{2g}$-$e_g$ transitions depend on the polarization of the excitation light with respect to the crystalline axes of a rutile $TiO_2(110)$ sample [48]. In other words, for polarized light excitation, rotating the $TiO_2$ crystal azimuth or the laser polarization selects between the specific intermediate $e_g$ states that can be excited by a resonant 2PP process. Temperature dependence of the $t_{2g}$-$e_g$ transitions recorded in 2PP spectra provides information on $e$-$p$ interactions affecting individually the Ti-3$d$ $t_{2g}$ and $e_g$ states. Moreover, we point out that photoemission measures $k$-resolved energies of the optically coupled bands relative to $E_F$, rather than their $k$-integrated light-absorption frequencies, such as are recorded in optical spectra. The resonant frequencies can as well be obtained by tuning the excitation photon energy $h\nu$ and recording the 2PP spectral intensity.

To explore the $e$-$p$ interaction in $TiO_2$, we record 2PP spectra of $TiO_2(110)$ employing the one-photon resonant $t_{2g}$-$e_g$ transition as function of $h\nu$ and sample temperature [47,48]. Photoemission is recorded with a Specs Phoibos 100 hemispherical electron energy analyzer. The tunable excitation light is generated by a Clark MXR Impulse fiber laser oscillator amplifier, which excites a noncollinear optical parametric amplifier at a 1-MHz repetition rate. Figure 1 shows the normalized, wavelength ($h\nu = 3.44$–3.54 eV)-dependent 2PP spectra of $TiO_2$ at 300 K for excitation with $p$-polarized light and the sample azimuth aligned with the $[1\bar{1}0]$ crystalline axis in the optical plane. Such measurements locate the absorption maximum based on intensity of the $t_{2g}$-$e_g$ transition. The spectra are normalized at the work function edge, where the signal is not resonance enhanced. The same transition can be accessed if the crystal is rotated so that the [001] axis is in the optical plane, and the excitation laser is $s$ polarized. The wavelength-dependent spectra in Fig. 1 and their analysis with respect to the observed $e_g$-state peak energy vs $h\nu$ in for the [001] direction in Fig. 2 show a dominant peak due to the $t_{2g}$-$e_g$ transition, which moves to a higher two-photon final-state photoelectron energy ($E_{fin}$) with increasing $h\nu$, and reaches maximum intensity at the $t_{2g}$-$e_g$ resonance. Performing similar measurements to Fig. 1 at other sample temperatures, as in Fig. 2, changes the $t_{2g}$-$e_g$ resonance energy on account of the $e$-$p$ interaction. Specifically, the resonance energy decreases from 3.61 to 3.44 eV as the sample temperature is increased from 31 to 600 K. To determine the temperature-dependent energies of the coupled $d$-orbital-derived bands, we plot $E_{fin}$ vs $h\nu$ in Fig. 2 for four different sample temperatures. In such plots: (1) the slope defines how many photons are necessary to coherently induce photoemission from a particular (initial or intermediate) state [86,87]; and (2) the intercept gives the energy from which the coherent excitation occurs. Because the slopes of such plots are ~1, the temperature-dependent intercepts tell us that photoemission occurs by absorption of a single photon from a transiently populated intermediate $e_g$ state and gives the temperature dependence of its energy. Because we also determine the $t_{2g}$-$e_g$ transition resonance energy at each temperature, we can also separately determine both the $t_{2g}$ and $e_g$-state energies at the $\Gamma$ point relative to $E_F$ at each temperature, and report the results in Fig. 3. This plot shows that both the $t_{2g}$ and $e_g$-state energies increase with temperature, but the $t_{2g}$ increase is larger causing the resonance energy to decrease with increasing temperature. These 2PP measurements provide experimental data for comparison with the calculated $t_{2g}$ and $e_g$ band shifts due to $e$-$p$ interaction.

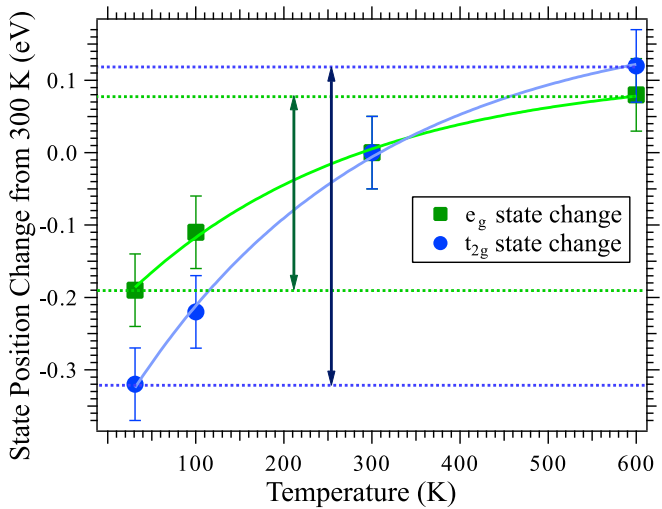

FIG. 3. Experimental change of the $e_g$ and $t_{2g}$ state energies with temperature from their reference positions at 300 K. The vertical arrows and dotted lines indicate the total shifts for each band in the measured temperature range.

Our goal then is to explain the temperature-dependent energy shifts of the $t_{2g}$ and $e_g$ conduction bands of TiO$_2$ from 2PP experiments by an *ab initio* description of the $e$-$p$ interaction; finally, based on these results, we aim to define general principles for temperature-induced band shifts for bands derived from the $d$ orbitals. We will show that the coupling between Ti-$3d$ orbitals and phonons of acoustic and optical branches is the key to explaining the temperature-induced band-energy shifts, and in principle, the theory can be applied to other metal oxides with similar electronic structures, such as Fe$_2$O$_3$ [73–75].

## IV. Ab Initio CALCULATIONS

Calculations are carried out with the all-electron, full-potential, numerical atomic orbitals-based code FHI-AIMS [88–95], whereby the finite-difference method is used to determine phonon frequencies and $e$-$p$ couplings. Polar corrections, which are necessary due to long-range electrostatic interactions [32,38,68,69], are applied using the method proposed by Nery and Allen [62]. For TiO$_2$, this correction only leads to a nearly constant shift in the temperature-dependence curves, as discussed in more detail, as well as the **q**-point convergence, in the Supplemental Material [63].

The primitive TiO$_2$ cell contains 6 atoms, which contribute 18 vibrational modes at every **q** point. We use a local density approximation for exchange and correlation (LDA parametrization of the correlation energy density of the homogeneous electron gas by Perdew and Zunger [96] based on the data of Ceperley and Alder [97]). We also calculate the zero-point renormalization using the HSE06 functional, and find that, compared to the LDA functional, the difference in renormalization energy of $t_{2g}$-$e_g$ splitting is about 1 meV, as shown in the Supplemental Material [63]. A "tight" default setting is used for basis sets and integration grids: the primitive cell is optimized until the force on each atom is smaller than $<10^{-5}$ eV/Å, and the obtained lattice constants for the primitive cell are $a = 4.549$ and $c = 2.919$ Å, which is in good agreement with previous calculations [98,99] using the LDA functional, and within 1% of the experimental values [100] of $a = 4.59$ and $c = 2.95$ Å. Furthermore, the electronic sampling of the Brillouin zone is performed with an $8 \times 8 \times 12$ **k** grid, and the phonon as well as the renormalizations by the $e$-$p$ interaction are performed with a $4 \times 4 \times 6$ **q** grid. The high-frequency dielectric tensor and Born effective charges to calculate the splitting between the transverse and longitudinal optical phonon modes are taken from Ref. [101].

This approach yields the temperature dependences of the VBM and the CBM, and therefore that of the electronic band gaps of both the ionic rutile TiO$_2$ and covalent SiO$_2$ materials, as shown in Fig. 4(b). In both cases, the VBMs involve the O-$2p$ orbitals; which monotonically increase in energy with temperature in our calculations. By contrast, the temperature dependences of CBMs of TiO$_2$ and SiO$_2$ have the opposite signs, which we attribute to their dominant orbital characters. More specifically, the energy of Ti-$3d$ $t_{2g}$ orbitals increases with temperature, whereas for the Si-$2s$ orbital of SiO$_2$ it decreases. These different temperature dependences of the CBMs of TiO$_2$ and SiO$_2$ confirm the hypsochromic shift of the former and the bathochromic shift for the latter. We note that two recent theoretical studies of TiO$_2$ by Wu *et al.* [102] and Monserrat [103] found an increase of the gap up to 300 K, followed by a decrease with increasing $T$ [102,103]. By making a careful comparison, we found the differences between Wu's data [102] and our calculations can be attributed to the different phonon frequencies as shown in Ref. [63] as well as different $e$-$p$ factors for the CBM state. These differences can be attributed to the pseudopotential method adopted by Wu *et al.* [102] and Monserrat [103], whereas we employ an all-electron full-potential method.

The opposite temperature dependences of the TiO$_2$ and SiO$_2$ CBMs can be explained by how the $e$-$p$ coupling affects energies of the dominant atomic orbitals. In a qualitative model based on simple phase-space arguments, one would expect the CBM to decrease and VBM to increase in energy with temperature, because their band curvatures [63], and therefore their band masses, have the opposite signs [104]. This rule indeed holds for SiO$_2$ and for the VBM of TiO$_2$, but not for its CBM, because it does not consider its dominant orbital character; a quantitative first-principles approach must also consider the sign and magnitude of the $e$-$p$ factor $f_{jq}$ for each mode $j$ and point $q$ summed over all modes to determine the slope of the temperature dependence of each band, as shown in Eq. (1). Consequently, as it is shown in Fig. 4(c), raising of the CBM energy with temperature in TiO$_2$ occurs because the number of modes with a positive $f_{jq}$ is larger than ones with a negative one. By contrast, the decrease of the CBM energy with temperature in SiO$_2$ occurs because all of $f_{jq}$ values are negative.

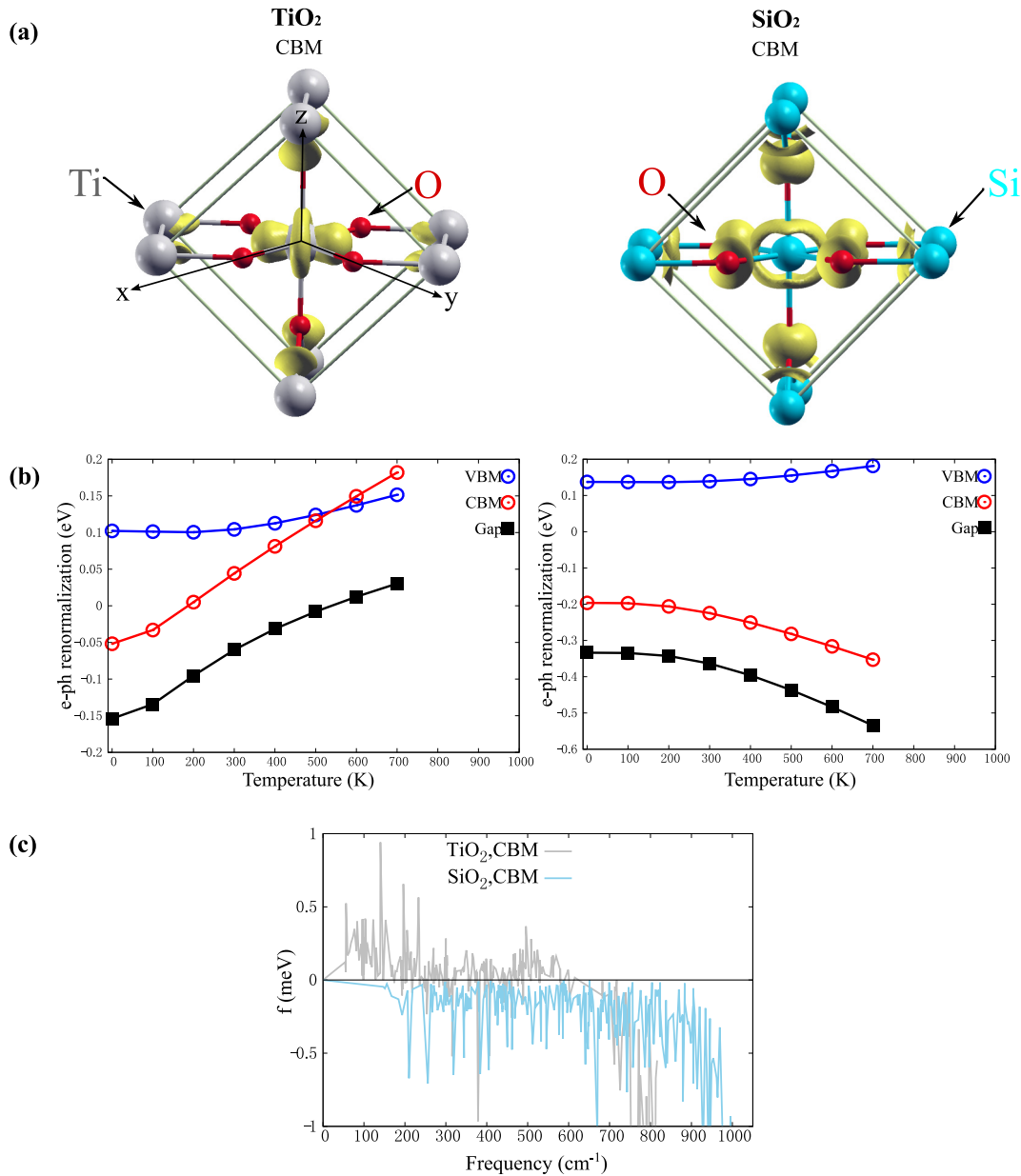

FIG. 4. (a) CBM orbitals of TiO$_2$ and SiO$_2$ are plotted within their unit cells (red balls-O ions, gray balls-Ti ions, blue balls-Si ions, yellow-CBM orbitals). (b) Calculated VBM- and CBM-state energy shifts for TiO$_2$ and SiO$_2$. The VBM shifts are caused by the O-2$p$ orbitals, and are similar for TiO$_2$ and SiO$_2$; for CBMs, the Ti-3$d$ and Si-2$s$ orbitals shift oppositely. Each type of orbital has a characteristic temperature-dependent behavior. (c) The $e$-$p$ factor $f_{jq}$ for phonons of increasing frequencies for the CBM states of TiO$_2$ (gray) and SiO$_2$ (blue), which cause their opposite shifts.

By comparing the calculations on rutile TiO$_2$ with stishovite SiO$_2$, their temperature-dependent band-energy renormalizations can be attributed to the different orbital characters of their CBMs. To confirm that Ti-3$d$ orbitals are responsible for the temperature upshift, we further investigate the temperature behavior of other 3$d$ electron orbitals in TiO$_2$. Specifically, we exploit the fact that by applying an octahedral crystal field on the $d$ orbitals the energy of the $t_{2g}$ states is reduced and that of the $e_g$ states is increased through the interaction with the ligands. We analyze how the $e$-$p$ interaction affects these orbitals differently and compare with the $t_{2g}$-$e_g$ spectra recorded in 2PP experiments to determine the $e$-$p$ renormalization of the $e_g$ states. Using Eqs. (1) and

(2), the temperature effects on the conduction band $t_{2g}$ and $e_g$ orbitals of rutile TiO$_2$ are calculated and shown in Fig. 5. One can see that as temperature increases from 0 to 600 K, both the energies of $t_{2g}$ and $e_g$ states increase, but we expect the $t_{2g}$-$e_g$ transition energy to decrease by around $-0.10$ eV through the $e$-$p$ effect; this decrease occurs because the increase of $t_{2g}$-state energies is larger than that for the $e_g$ states.

In addition to the above discussed atomic contribution to the band-energy renormalization by thermal phonon excitation, the lattice expansion makes an additional contribution that quantitatively affects the temperature dependence of the $t_{2g}$ to $e_g$ transition energy, and is likely to be important in other $d$-electron bands that are affected by crystal-field splitting and have anharmonic phonon modes. In our calculation of the temperature-dependent structure of $TiO_2$, the *a* and *c* lattice parameters are found to increase by 0.02 Å as the temperature is increased from 0 to 600 K, due to anharmonic interatomic potentials, in good agreement with the literature [105,106]. The lattice expansion does not, however, change the discussed qualitative behavior, but only causes the $t_{2g}$ to $e_g$ transition energy to shrink additionally by $-0.076$ eV, when temperature is increased from 0 to 600 K [63]. In other words, the lower $t_{2g}$ orbital energy increases, while the upper $e_g$ orbital energy decreases due to weaker interaction with the ligands that is caused by the lattice expansion. In total, the first-principles calculations predict a quantitative downshift of the *d*-*d* transition by $-0.176$ eV between 0 and 600 K due to the combined effects of *e*-*p* band-energy renormalization and lattice expansion, in excellent agreement with the experimental value of $-0.17$ eV, as shown in Fig. 6.

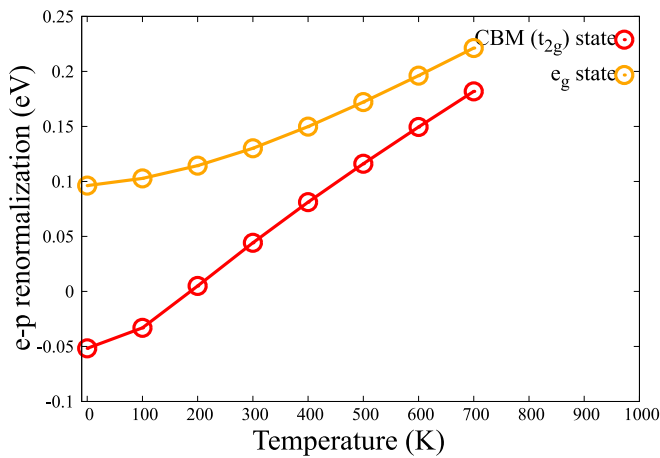

FIG. 5. Calculated temperature dependence of the *e*-*p* renormalization of the CBM ($t_{2g}$) and $e_g$ states.

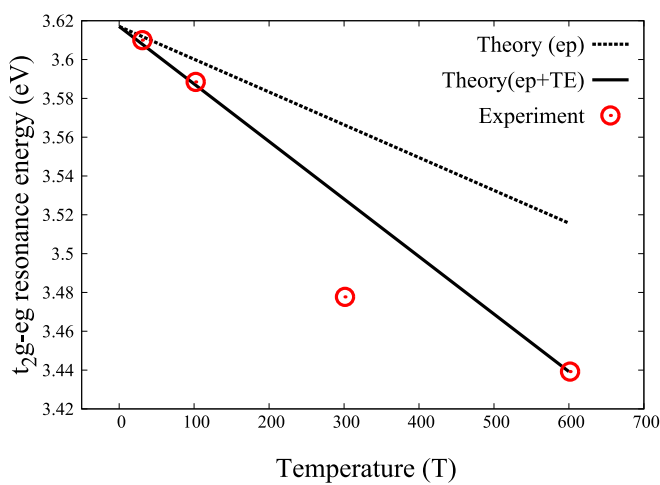

FIG. 6. Theoretical and experimental resonance energies for optical transitions from the $t_{2g}$ to $e_g$ states. The dotted line represents the theoretical prediction including only the *e*-*p* renormalization. The solid line represents the total theoretically predicted shift, which also includes the thermal expansion (TE). By considering both contributions, the theoretical value comes into excellent agreement with the experimental data (red points).

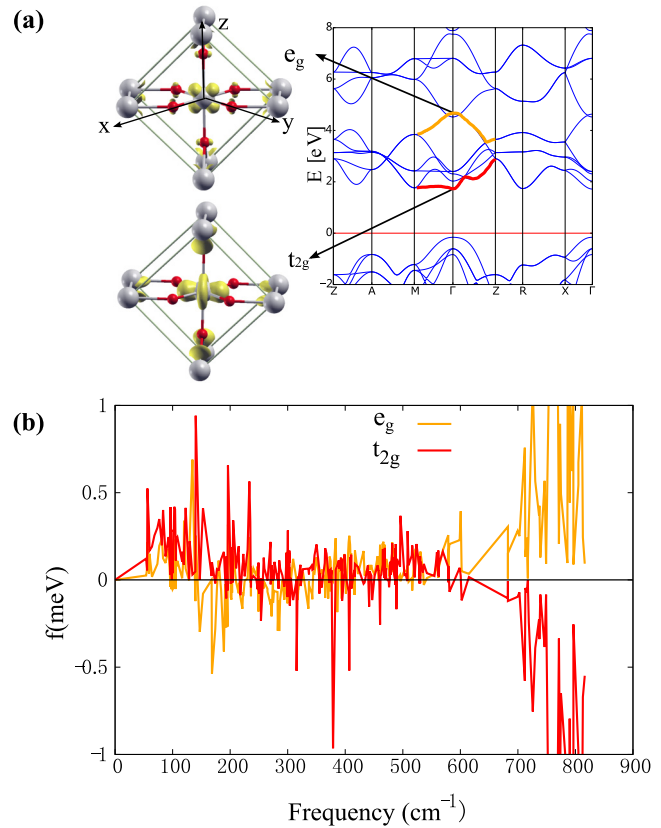

FIG. 7. (a) Electronic orbitals ($t_{2g}$ and $e_g$) of $TiO_2$ involved in the optical transitions that contribute to the 2PP spectra at the $\Gamma$ point. The $t_{2g}$ state has the $d_{xy} + d_{z^2}$ orbital character while the $e_g$ state has the $d_{xz} + d_{yz}$ orbital character [47,81]. (b) The *e*-*p* factor with respect to phonon frequencies for the $t_{2g}$ and $e_g$ states.

To analyze the influence of the specific symmetry *d* orbitals on *e*-*p* coupling, we further quantify the behavior of those with $t_{2g}$ and $e_g$ character in $TiO_2$. The $t_{2g}$ and $e_g$ orbitals derived from the conduction bands are plotted separately in Fig. 7(a). With careful analysis, we find the percentage of the positive *e*-*p* factors for the $t_{2g}$ (73%) is larger than the $e_g$ state (69%). Furthermore, it is clear from Fig. 7(b) that the most positive factors of the $t_{2g}$ state are in the low-energy part of the phonon spectrum, which has larger population factors than the high-energy part according to Bose-Einstein statistics, causing the temperature-dependent slope of upward renormalization to be steeper for the $t_{2g}$ bands as compared with the $e_g$ bands, as is evident in Fig. 5.

The observed *e*-*p* renormalization of the $t_{2g}$ and $e_g$ states in $TiO_2$ can be extended in general to other materials where the *d* orbitals experience a crystal field with the octahedral symmetry. The crystal field in metal oxides with tetragonal coordination, however, has the opposite crystal-field splitting such that the $e_g$ orbitals are stabilized and the $t_{2g}$ orbitals are destabilized, so that their relative energies are reversed, and thus we expect phonons to have the opposite effect on them than in the octahedral symmetry. Other metal coordination structures with different symmetries cause specific and well-understood splitting of the $d$ orbitals, from which it is possible to use similar calculations and analysis to predict how the $e$-$p$ interaction will renormalize their energies. Although we have not calculated any $f$-electron system, transition metals where $f$-electron orbitals participate in bonding and in optical spectra will be affected by crystal-field interactions in a similar fashion as the $d$-electron orbitals.

## V. CONCLUSIONS

In summary, we have shown that the dissimilar temperature dependences of electronic bands in $TiO_2$ and $SiO_2$ derive from how the crystal field modulates the interactions of $d$- and $s$ orbitals with the ligand ions. Although this type of analysis can rationalize the band symmetry-dependent trends, a more quantitative calculation of the $e$-$p$ interaction must consider the vibrational band-dependent $e$-$p$ coupling factor. Our systematic analysis establishes the connection between $d$-orbital interactions with their crystal field, which leads to the opposite temperature behavior for the $d$- and $s$-orbital-derived bands, as explicitly shown for the $TiO_2$ and $SiO_2$ model compounds. The calculated $e$-$p$ renormalization energies including the contributions from the $TiO_2$ lattice expansion are in excellent agreement with the measured $t_{2g}$ and $e_g$ band-energy shifts with temperature as measured by 2PP spectroscopy. We emphasize that the simple concepts presented here provide guidance on how to interpret and design $e$-$p$ interactions involving $d$-electron atoms to tailor the temperature dependence of electron and spin properties such as superconductivity and ferromagnetism. We conclude that the $d$-orbital-derived band-energy shifts with temperature, which may have been perceived to be as anomalous, are just the characteristic properties of $d$-electron-derived bands that are commonly found in other metal oxides or semiconductors. The specific shifts of $d$-orbital-derived bands will depend on the symmetry of the ligand field, which causes the otherwise degenerate $d$ orbitals to split due to the attractive, neutral, or repulsive interactions with the surrounding ligands. For example, the crystal-field-dependent differences in the temperature renormalizations of $d$-orbital bands could be responsible for electronic phase transitions such as occurs in $VO_2$ [107]. We expect that the physical concepts for temperature-dependent band shifts can be applied broadly to qualitatively understand the temperature-dependent electronic band structures of other materials based on the nature of the atomic orbitals that form the bands and the crystal symmetry that tunes their energies, without a need to perform high-level theoretical calculations.

### ACKNOWLEDGMENTS

We acknowledge funding from NSF Grant No. CHE-1565842. H.P. acknowledges support from the Alexander von Humboldt Foundation, the Chinese Academy of Sciences President's International Fellowship Initiative, and Luo Jia Visiting Professorship of the Wuhan University. H.S. acknowledges the funding from the Strategic Priority Research Program of Chinese Academy of Science (Grant No. XDC01040100). H.S. and C.C. acknowledge the funding from the Einstein Foundation (project ETERNAL), the Deutsche Forschungsgemeinschaft (DFG) through Grant No. SFB 951, and the European Union Horizon 2020 research and innovation program under Grant Agreement No. 676580 with The Novel Materials Discovery (NOMAD) Laboratory, a European Center of Excellence. P.R. acknowledges financial support from the Academy of Finland through its Centers of Excellence Program (Projects No. 251748 and No. 284621).

H.S. was primarily responsible for the theoretical calculations and A.A. was primarily responsible for the experimental temperature-dependence study of electronic bands of $TiO_2$.